\documentclass[runningheads]{llncs}
\usepackage{floatrow}
\usepackage{cite}
\usepackage{graphicx}
\usepackage[dvipsnames]{xcolor}
\usepackage{caption}
\usepackage{subcaption}
\usepackage{amsmath}
\usepackage{pifont}
\newcommand{\cmark}{\ding{51}}%
\newcommand{\xmark}{\ding{55}}%

\usepackage{array}
\newcolumntype{L}[1]{>{\raggedright\let\newline\\\arraybackslash\hspace{0pt}}m{#1}}
\newcolumntype{C}[1]{>{\centering\let\newline\\\arraybackslash\hspace{0pt}}m{#1}}
\newcolumntype{R}[1]{>{\raggedleft\let\newline\\\arraybackslash\hspace{0pt}}m{#1}}

\begin{document}
\title{Learning normal asymmetry representations for~homologous~brain~structures}
%
\titlerunning{Learning normal asymmetry representations for homologous brain structures}
%

\author{Duilio~Deangeli\inst{1,2}
\and
Emmanuel~Iarussi\inst{2,3}
\and
Juan~Pablo~Princich~\inst{4}
\and
Mariana~Bendersky~\inst{3,4}
\and
Ignacio~Larrabide\inst{1,2}
\and
José~Ignacio~Orlando\inst{1,2}
}
\authorrunning{D. Deangeli et al.}

\institute{Yatiris, PLADEMA, UNICEN, Tandil, Buenos Aires, Argentina \and
CONICET, Buenos Aires, Argentina \and
Universidad Torcuato Di Tella, CABA, Argentina \and
ENyS, CONICET-HEC-UNAJ, Florencio Varela, Buenos Aires, Argentina \and
Normal Anatomy Department, UBA, CABA, Argentina
}


\maketitle              
\begin{abstract}
Although normal homologous brain structures are approximately symmetrical by definition, they also have shape differences due to e.g. natural ageing. On the other hand, neurodegenerative conditions induce their own changes in this asymmetry, making them more pronounced or altering their location. Identifying when these alterations are due to a pathological deterioration is still challenging. Current clinical tools rely either on subjective evaluations, basic volume measurements or disease-specific deep learning models. This paper introduces a novel method to learn normal asymmetry patterns in homologous brain structures based on anomaly detection and representation learning. Our framework uses a Siamese architecture to map 3D segmentations of left and right hemispherical sides of a brain structure to a normal asymmetry embedding space, learned using a support vector data description objective. Being trained using healthy samples only, it can quantify deviations-from-normal-asymmetry patterns in unseen samples by measuring the distance of their embeddings to the center of the learned normal space. We demonstrate in public and in-house sets that our method can accurately characterize normal asymmetries and detect pathological alterations due to Alzheimer's disease and hippocampal sclerosis, even though no diseased cases were accessed for training. Our source code is available at https://github.com/duiliod/DeepNORHA.
\end{abstract}

\section{Introduction} 
\label{sec:introduction}


(Sub)cortical brain structures are approximately symmetrical between the left and right hemispheres~\cite{tortora2018principles}. 
Although their appearance and size are similar, they usually present difficult-to-characterize morphometric differences that vary among healthy populations~\cite{woolard2012anatomical}, e.g. due to natural ageing~\cite{ardekani2019sexual}.
Moreover, it has been studied that some neurological conditions, including Alzheimer's disease~\cite{andrade2015defining, herzog2021brain} (AD), schizophrenia~\cite{csernansky2004abnormalities}, and epilepsy~\cite{bernasconi2003mesial,park2022topographic}, are associated to asymmetry of the hippocampus or the amygdala~\cite{wachinger2016whole}.
In regular medical practice, radiologists detect pathological changes in asymmetry by manually inspecting the structure using brain MRIs~\cite{bernasconi2003mesial}.
They rely on their own subjective experience and knowledge, which varies among observers and lacks reproducibility, or standard quantitative measurements, e.g. volume differences in segmentations~\cite{andrade2015defining,herbert2005brain, princich2021diagnostic, ardekani2019sexual,liu2019using,park2022topographic}, which fail to capture morphological asymmetries beyond differences in size~\cite{richards2020increased,herzog2021brain}. 
No tools automate quantifying normal asymmetry patterns beyond volume~\cite{andrade2015defining}, e.g. to detect deviations caused by a neurodegenerative disease. Some deep learning tools approximate this goal in a binary classification setting to differentiate one particular condition from normal cases~\cite{fu2021altered,borchert2021artificial}. However, this form is heavily specialized to discern asymmetry alterations associated with one specific disease, requiring retraining for every new condition~\cite{li2021hippocampal,liu2019using}.

This paper introduces a novel framework for learning NORmal Asymmetries of Homologous cerebral structures (deep NORAH) based on anomaly detection and representation learning. 
Unlike previous methods that train Siamese neural networks with volume descriptors~\cite{liu2019using}, our model takes 3D segmentations of left and right components from MRIs as inputs, and maps them into an embedding that summarizes their shape differences. 
Essentially, our Siamese architecture includes a shape characterization encoder that extracts morphological features directly from segmentations and an Asymmetry Projection Head (APH) that merges their differences to create a compact representation of asymmetries.
To ensure this embedding learns the heterogeneity in normal individuals, our network is trained only with healthy samples, using a self-supervised pre-training stage based on a Contractive Autoencoder (CAE) and then fine-tuning using a Support Vector Data Description (SVDD) objective. Our experiments in the hippocampus show that our model can easily project new cases to the normal asymmetry space. Furthermore, we show that the distance between the embedding and the center of the normal space is a measure of deviation-from-normal-asymmetry, as we empirically observed increased distance in pathological cases. Hence, deep NORAH can be used to diagnose, e.g. AD, hippocampal sclerosis and even mild cognitive impairment (MCI) by simply detecting the differences in asymmetry regarding the normal set, without needing diseased cases for training. 

In summary, our contributions are as follows: (i) ours is the first unsupervised deep learning model explicitly designed to learn normal asymmetries in homologous brain structures; (ii) although it is trained only with normal data, it can be used to detect diseased samples by quantifying the degree of deviation with respect to a healthy population
, unlike existing methods that capture only disease-specific asymmetries~\cite{li2021hippocampal}; and (iii) compared to other state-of-the-art anomaly detection approaches, our method demonstrates consistently better results for discriminating both synthetic and diseased-related asymmetries.

\section{Methods} 
\label{sec:methods}

\begin{figure}[t]
\centering
\includegraphics[width=0.65\textwidth]{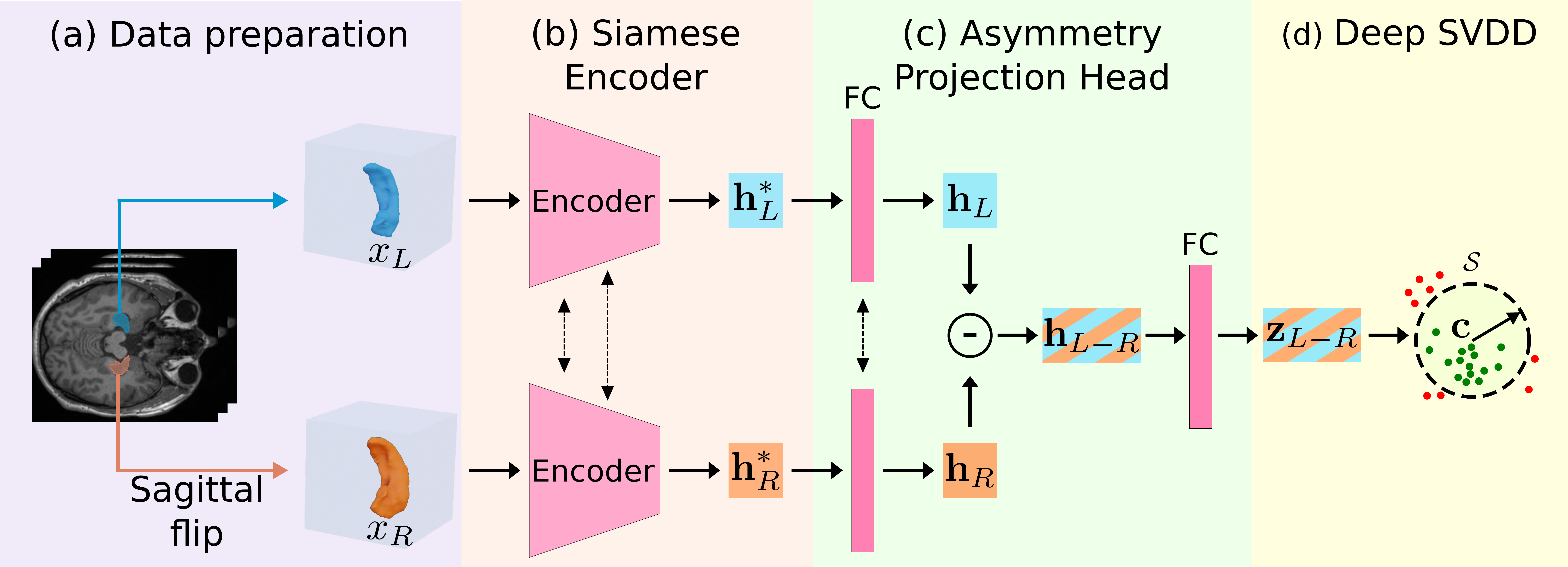}
\caption{Schematic of our framework for capturing abnormal asymmetries in homologous brain structures from 3D MRI.}
\label{fig:schematic-test}
\end{figure}

Fig.~\ref{fig:schematic-test} depicts a flowchart of our method as applied in test time. Our goal is to automatically measure the asymmetry of a given homologous brain structure $x = (x_L, x_R)$, with $x_L$ and $x_R$ being the 3D segmentations of its left and right lateral elements, and learn if these differences are typical for a normal population. To do so, we propose to learn a Siamese deep neural network $\mathbf{F}_\theta(x) = \mathbf{z}_{L-R}$ with $\theta$ parameters using an anomaly detection objective and samples from healthy individuals. $\mathbf{z}_{L-R}$ is a compact representation of the asymmetries in $x$, obtained by learning a hypersphere $\mathcal{S}$ with a center $\mathbf{c}$ and minimum radius. In test time, samples with normal asymmetries are projected to the vicinity of $\mathbf{c}$, while those with unexpected differences fall far from this point. As a result, the distance $d = {\| \mathbf{F}_\theta(x) - \mathbf{c} \|}^2_2$ can be used as a deviation-from-normal-asymmetry index.


To train this model, we first learn a 3D shape encoder $f_{\theta_{\text{EN}}}(x)$ as part of a CAE, using normal samples (Section~\ref{subsec:siamese-encoder}). 
This network can take any single segmentation of a lateral element $x_{(i)}$ as input, and map it to a high dimensional shape representation $\mathbf{h}^{*}_{(i)}$.
We then add this encoder to a Siamese architecture by attaching it to an APH, which captures the differences in shape from $\mathbf{h}^{*}_{L}$ and $\mathbf{h}^{*}_{R}$, and project them into the unique asymmetry embedding $\mathbf{z}_{L-R}$.
To this end, both the pre-trained encoder and the APH are trained using a deep SVDD objective (Section~\ref{subsubsec:oc-svdd}).
This second learning phase not only trains the APH from scratch but fine-tunes the shape encoder to capture those morphological characteristics that are the most common source of asymmetry in normal individuals.

\subsection{Pre-training the shape characterization encoder as a CAE}
\label{subsec:siamese-encoder}

\begin{figure}[t!]
     \centering
     
     \begin{subfigure}{0.35\textwidth}
         \centering
         \includegraphics[width=\textwidth]{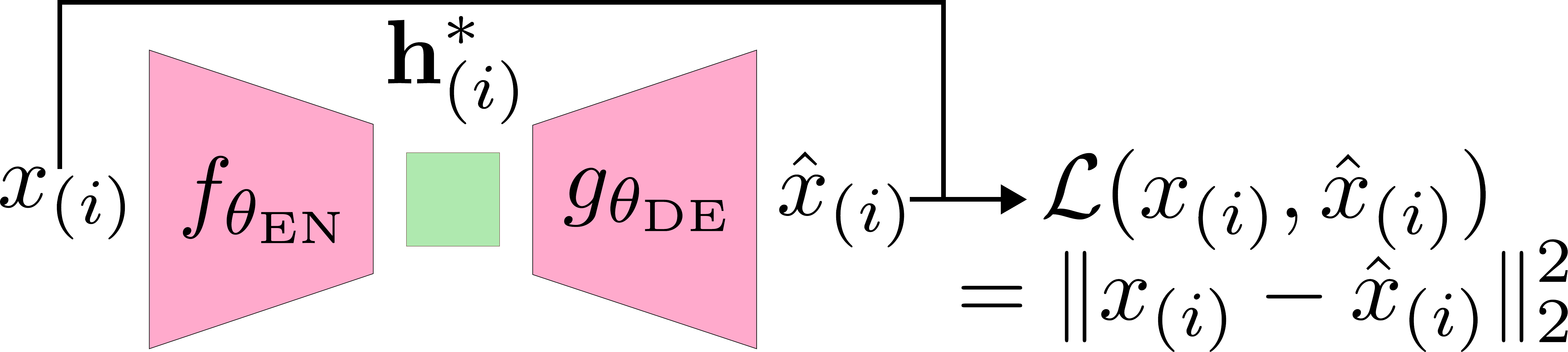}
         \caption{Pre-training}
         \label{fig:AE_BASED_PRETRAIN}
     \end{subfigure}
     \quad
     \begin{subfigure}{0.35\textwidth}
         \centering
         \includegraphics[width=\textwidth]{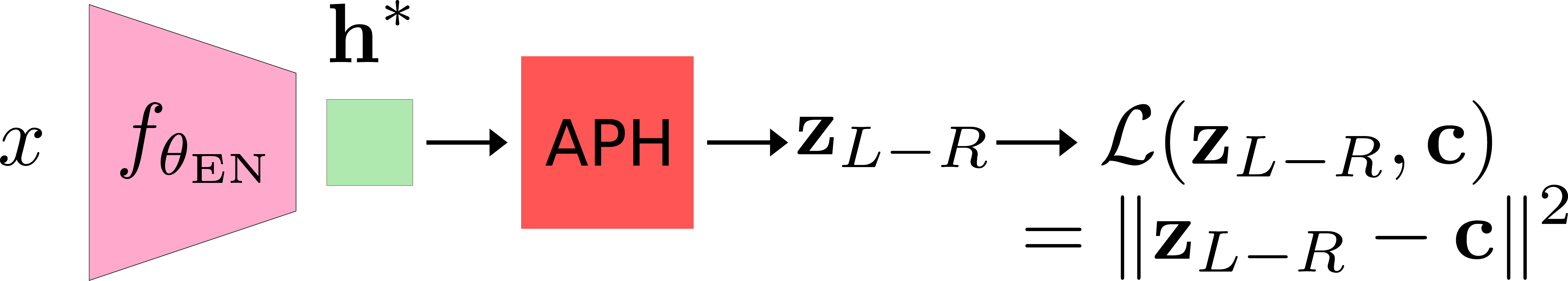}
         \caption{Fine-tuning}
         \label{fig:Deep_SVDD_FINE_TUN}
     \end{subfigure}
     \caption{Training sequence of our method.}
\end{figure}

Our shape encoder $f_{\theta_{\text{EN}}}$ indistinctly map an arbitrarily left or right segmentation $\mathbf{x}_{(i)}$ of an homologous structure $x$, to a feature vector $\mathbf{h}^{*}_{(i)}$ that describes its shape. To this end, we apply a warm-up learning phase that trains $f_{\theta_{\text{EN}}}$ as the encoding path of a CAE, using a self-supervised learning loss (Fig.~\ref{fig:AE_BASED_PRETRAIN}). Hence, the encoder is simultaneously trained with a decoder $g_{\theta_{\text{DE}}}(i)$ using a reconstruction task. The encoder compresses the input into a lower-dimensional representation $\mathbf{h}^{*}_{(i)}$ by applying a series of convolutional and pooling operations, and the decoder tries to reconstruct it using upsampling operations and convolutions.

Formally, let $(g_{\theta_{\text{DE}}} \circ f_{\theta_{\text{EN}}})(x_{(i)})$ be a convolutional CAE with a decoding path $g_{\theta_{\text{DE}}}(\mathbf{h}^{*}_{(i)})$ with parameters $\theta_{DE}$ that outputs a reconstruction $\hat{x}_{(i)}$ of the input $x_{(i)}$ from its hidden representation $\mathbf{h}^{*}_{(i)}$. This is achieved by minimizing a mean square error objective (Fig.~\ref{fig:AE_BASED_PRETRAIN}). After it, the decoder is discarded and the shape encoder is used in the Siamese setting of our anomaly detection network.

\subsection{Learning normal asymmetries with a Siamese network}
\label{subsec:projection-head}




\subsubsection{Assymetry projection head.} The purpose of our APH is to project the shape representation $\mathbf{h}^{*}$ obtained by $f_{\theta_\text{EN}}$ into a compact embedding $\mathbf{z}_{L-R}$ (Fig.~\ref{fig:Deep_SVDD_FINE_TUN}) that better describes normal population asymmetry characteristics. In our implementation, this network is a multilayer perceptron (MLP) with two fully connected (FC) layers separated with a ReLU activation. The first FC layer is used in a Siamese setting by feeding it with the shape representations $\mathbf{h}^{*}_{L}$ and $\mathbf{h}^{*}_{R}$ of the left and right elements, respectively. Each of these inputs are projected into two new feature vectors $\mathbf{h}_{L}$ and $\mathbf{h}_{R}$ with a lower dimensionality. A merging operation (e.g. subtraction or concatenation) combines them into a joint representation $\mathbf{h}_{(L-R)}$, which is projected by the second FC layer into the asymmetry embedding $\mathbf{z}_{(L-R)}$. Notice that the main design choices to be made are the dimensionality of the outputs of each FC layer and the merging operation.




\subsubsection{One-Class Deep SVDD.}\label{subsubsec:oc-svdd} In order to enforce $\mathbf{z}_{(L-R)}$ to represent the asymmetry characteristics of normal individuals, we train the Siamese architecture $\mathbf{F}_\theta(x)$ in Fig.~\ref{fig:Deep_SVDD_FINE_TUN} using an anomaly detection objective. We adopted the one-class deep SVDD approach proposed in~\cite{ruff2018deep}, which solves:
\begin{equation}
     \min_\theta \frac{1}{n} \sum_{i=1}^{n} \| \mathbf{F}_\theta(x_{(i)}) - \mathbf{c} \|^{2} + \frac{\lambda}{2} {\| \theta \|}^2_2.
    \label{eq:svdd}
\end{equation}
The first term in Eq.~\ref{eq:svdd} is a quadratic loss that penalizes the Euclidean distance of $\mathbf{z}_{(L-R)}$ from the center $\mathbf{c}$ 
of a hypersphere $\mathcal{S}$, that is implicitly determined by the distance itself in the representation space. The second term is a classic weight decay regularizer, controlled by $\lambda$. Notice that we do not contract $\mathcal{S}$ by explicitly penalizing its radius and samples lying outside its boundary, but by minimizing their mean Euclidean distance with respect to $\mathbf{c}$~\cite{ruff2018deep}. To avoid convergence to a collapsed trivial solution with all zero weights, all layers in $\mathbf{F}_\theta$ do not use bias terms~\cite{ruff2018deep}, and the center $\mathbf{c}$ was set to the average $\mathbf{z}_{(L-R)}$ obtained by feeding $\mathbf{F}_\theta$ with all training samples before fine-tuning, as in~\cite{zhang2021anomaly}. 
At that stage, $f$ is already pre-trained using the self-supervised strategy described in Section~\ref{subsec:siamese-encoder}, but the APH has random weights. Nevertheless, we experimentally observed that this center $\mathbf{c}$ is already enough to avoid a collapsed $\mathcal{S}$.

\subsubsection{Deviation-from-normal-asymmetry index.} 
The distance between the asymmetry embedding of an unseen sample $x$ and the center of the learned hypersphere, $s(x;\mathbf{c}) = \| \mathbf{F}_\theta(x) - \mathbf{c} \|^{2}$, can be used as a deviation-from-normal-asymmetry index: when the input $x$ is a normal sample, its asymmetry embedding is expected to lie in the vicinity of $\mathbf{c}$, then associated to a small $s$ value; on the other hand, if $x$ is the segmentation of a subject with abnormal asymmetries, its associated $\mathbf{z}_{(L-R)}$ will lie afar from $\mathbf{c}$, reporting a higher $s$ value.

  


\section{Experimental setup} 
\label{sec:experiments}


We studied our method for hippocampal asymmetry characterization as a use case.
First, we tested its ability to capture deviations in asymmetry using synthetically altered hippocampi with increased deformations, in a controlled setting. 
Then, we indirectly evaluated its performance as a diagnostic tool for  neurodegenerative conditions, using $s$ as a deviation-from-normal-asymmetry index. 
Finally, we performed an ablation study to understand the influence of design factors such as the shape encoder architecture, APH size, and merging operation. 

\subsubsection{Materials}
\label{subsec:materials}

We used a total of 3243 3D T1 brain MRIs, including 2945 from normal control (NC) individuals, 71 from patients with MCI, 179 with AD, and 16 and 32 with right (HSR) and left (HSL) hippocampal sclerosis, respectively. 
Samples were retrospectively collected from OASIS~\cite{marcus2010open} (NC = 2217, AD = 33), IXI~\cite{IXI} (NC = 539) and ADNI~\cite{ADNI} (NC = 53, AD = 33, MCI = 71) public sets, and from two in-house databases, ROFFO (NC = 83) and HEC (NC = 53, HSL = 32, HSR = 16), (see supp. mat. for demographics characteristics). 
All images were integrated in a single set, that we split into training, validation and test. 
The training set was used to learn patterns of normal asymmetry, with NC from ROFFO (63), IXI (539), and 70\% of the NC from OASIS. Ages ranged from 19 to 95 years old, to ensure capturing normal variations due to natural aging. 
60 NC images were kept aside for validation (see below).
The test sets, were used to evaluate the diagnostic ability of our method on different cohorts. 
TEST-ADNI and TEST-HEC sets include all subjects from ADNI and HEC sets, while TEST-OASIS includes all AD and the remaining 30\% of NC from OASIS. 

Images from different devices were aligned and normalized to a standard reference MNI T1 template using SPM12~\cite{penny2011statistical}. Hippocampal segmentations were obtained using HippMapp3r~\cite{goubran2020hippocampal}, a CNN model that is robust to atrophies and lesions. The resulting segmentations were cropped to create separate masks for each hippocampus, each with size $64\times64\times64$ voxels.
We created synthetic validation sets with abnormal hippocampal asymmetries to study the model's ability to identify asymmetries of different variations. We used 60 normal hippocampal segmentations (20 from ROFFO and 40 from OASIS) and applied elastic deformations~\cite{van2021elasticdeform} with $\sigma \in \{ 3, 5, 8 \}$ to one of the two hippocampi of 20 individuals, resulting in 60 simulated abnormal pairs (see supp. mat. for qualitative examples). Four validation sets $S_\sigma$ were created, with $\sigma=3, 5, 8$ and all, including the original 60 normal pairs and the corresponding simulated cases in the first three and a mix of all of them in the last one. This allowed us to perform hyperparameter tuning and evaluate the model's performance.

\subsubsection{Implementation details.}
We studied two backbones for our Siamese shape encoder: a LeNet-based architecture similar to the one in~\cite{ruff2018deep} (but adapted to 3D inputs), with 3 convolutional layers with 16, 32 and 64 $5 \times 5 \times 5$ filters, each followed by a 3D batch normalization (BN) layer and a ReLU operation;  
and a deeper CNN equal to the encoding path of the CAE in~\cite{oktay2017anatomically} to learn anatomical shape variations from 3D organs. The size of the FC layers in the APH were adjusted based on the validation set. For CAE pre-training, the LeNet based encoder was attached to a bottleneck FC layer, followed by a symmetric decoder with 3 trilinear interpolation upsamplings, each followed by a convolutional layer with 3D BN and ReLU. For the deeper encoder, on the other hand, we used the exact same CAE from~\cite{oktay2017anatomically}. Further details about the architectures are provided in the supp. mat. We pre-trained the networks using the CAE approach for 250 epochs, and then fine-tuned them with SVDD for another 250 epochs. In all cases, we used Adam optimization with a learning rate of $10^{-4}$, weight decay regularization with a factor of $10^{-6}$ and a batch size of 12 hippocampi pairs. We used PyTorch 1.12 and SciKit Learn for our experiments.

\subsubsection{Baselines.}
We compared our model with respect to other multiple approaches. 
To account for the standard clinical methods, we included the absolute and normalized volume differences (AVD and NVD, respectively), used in~\cite{pedraza2004asymmetry} as scores for asymmetry.
We also included shallow one-class support vector machines (OC-SVMs)~\cite{scholkopf1999support} trained with the same NC subjects than ours but different feature sets.
We used ShapeDNA~\cite{wachinger2016whole} (ShapeDNA + OC-SVM), which was previously studied to characterize hippocampal asymmetries~\cite{richards2020increased}, and a combined large feature vector (LFV + OC-SVM) including volumetric differences, ShapeDNA and shape features obtained using PyRadiomics (sphericity, compactness, quadratic compactness, elongation, flatness, spherical disproportion, surface volume ratio, maximum 2D diameter, maximum 3D diameter and Major Axis). 
For standard deep practices in anomaly detection, we trained hybrid CAE + shallow OC-SVM using our LeNet (LeNet-CAE + OC-SVM) and deeper backbones (Oktay \textit{et al.}~\cite{oktay2017anatomically} + OC-SVM).
Finally, a binary network was trained to detect AD cases (AD classification), in order to have a supervised counterpart for comparison. We used a larger training set that, apart from the same NC subjects used for the anomaly detection models, included all samples in TEST-OASIS. The validation set had in this case the remaining NCs from OASIS and AD cases from ADNI. The same backbone architecture was used, but with an additional FC layer that had softmax activation for classification, as in~\cite{li2021hippocampal}.

\section{Results \& Discussion}

\subsection{Characterization of normal and disease related asymmetries} 
\label{subsec:results-characterization}

\begin{figure}[t!]
\centering
\includegraphics[width=0.85\textwidth]{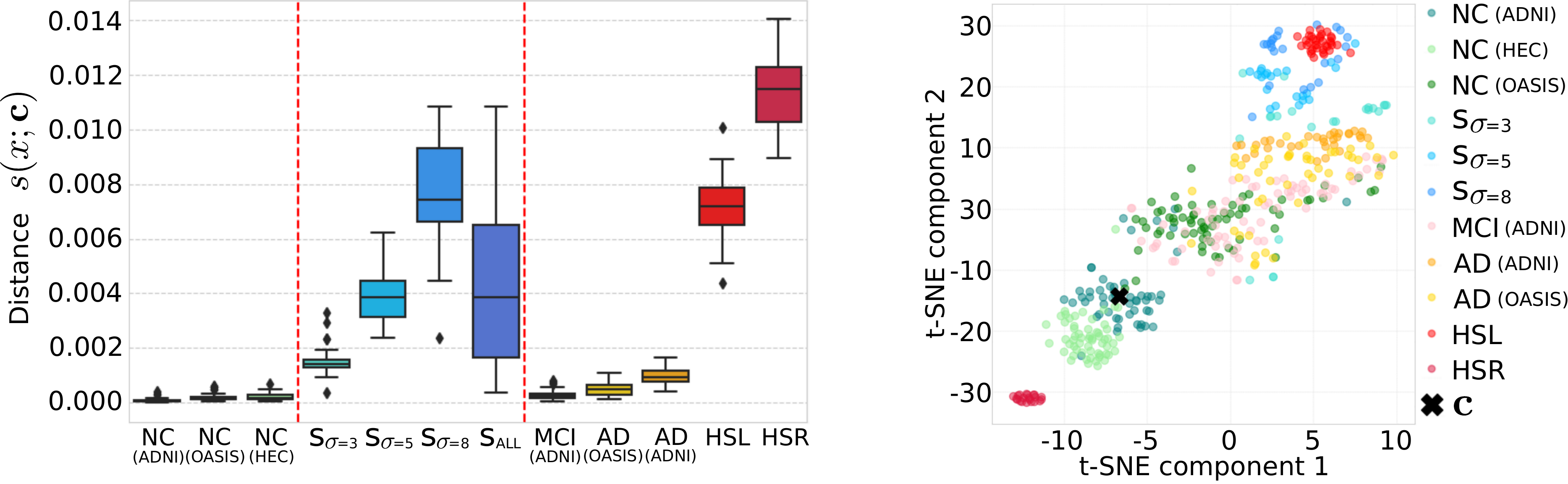}
\caption{Evaluation of our approach for characterizing different asymmetry patterns. Left: Distribution of distances $s(x;\mathbf{c})$ for each disease and NC group. Right: $t$-SNE representation of the normal asymmetry embeddings.}
\label{fig:boxplot_n_tsne}
\end{figure}

To test our hypothesis that samples with abnormal hippocampal asymmetries deviate from the center of the normal hypersphere, we evaluated the distances $s(x;\mathbf{c})$ between all samples in the validation and test sets and grouped them by disease category. The distribution of these distances is shown in Fig.~\ref{fig:boxplot_n_tsne} (left), with statistical significance assessed using Mann-Whitney-Wilcoxon rank-sum tests ($\alpha = 0.05$) with Bonferroni correction applied for multiple comparisons. 
Fig.~\ref{fig:boxplot_n_tsne} (right) represent the $t$-SNE projection of all representations. NC subjects are closely grouped in this plot, with the smallest distances to the center among all groups. These values are significantly lower than those obtained for synthetically altered samples ($p < 0.004$) and individuals with MCI ($p < 0.017$), AD ($p < 0.0083$), HSL ($p < 0.017$) and HSR ($p < 0.017$). Distances increase proportionally to $\sigma$ for synthetic cases and conditions with unilateral hippocampus shrinkage such as HSL and HSR, which are located at the extremes in the t-SNE representation. Synthetic cases with $\sigma=8$ group around one of the clusters, while those with $\sigma=5$ and 3 are scattered closer but still far from the center. MCI subjects from ADNI are scattered similarly to NC samples from OASIS, which is consistent with their distances. This could be due to cognitive decline in MCI cases not necessarily associated with alterations in hippocampal asymmetry, which can resemble that of NC, but rather with changes in other brain areas such as amygdala or thalamus~\cite{wachinger2016whole,low2019asymmetrical}. 
It is possible then that the MCI subjects in this study do not exhibit hippocampal asymmetry changes large enough to be distinguished from those in healthy controls. AD cases, on the other hand, are seen far from the hypersphere center as well, reporting distances higher than those from NC.
Finally, when compared one another, we observed significant differences in the distances between NC from ADNI set ($p < 0.0055$), but not between OASIS and HEC sets ($p < 0.1241$).

\subsection{Comparison with other approaches}
\label{subsec:results-comparison}

The evaluation results of different methods are displayed in Table~\ref{AUC table}. Volumetric based approaches only detected unilateral atrophies in synthetic and HSL/R cases and poor performance for MCI and AD. Feature-based methods performed slightly better for AD and synthetic but dropped for HSL/R and MCI, lacking the required robustness to abnormal asymmetry changes. Hybrid CAE + OC-SVM models had good performance on synthetic data but not on real data tasks. Finally, the AD binary classifier was only able to detect AD but failed in any other cases, which is consistent with the disease specialization hypothesis. Our method, on the contrary, outperformed all other methods for detecting both synthetically induced and pathological changes in hippocampal asymmetry. Notice, however, that the AD classification method is not a state-of-the-art approach but a comparable model with approximately the same backbone than ours. Other alternatives such as~\cite{liu2019using,herzog2021brain} might achieve much higher AUC values.

\begin{table}[t!]
      \centering
            \caption{AUC (95\% CI)  table for different approaches.}\label{AUC table}
\resizebox{0.9\textwidth}{!}{
            \begin{tabular}{c|C{2.3cm}|C{2.3cm}|C{2.3cm}|C{2.3cm}|C{2.3cm}|C{2.3cm}}
            \hline
            \textbf{Method}  &  \textbf{Synthetic} & \textbf{MCI} & \textbf{AD (ADNI)} & \textbf{AD (OASIS)} & \textbf{HSL} & \textbf{HSR}\\
            \hline
            AVD  & 0.72 {\scriptsize (0.62-0.81)} & 0.56 {\scriptsize (0.46-0.66)} & 0.55 {\scriptsize (0.42-0.67)} & 0.62 {\scriptsize (0.51-0.74)} & 0.95 {\scriptsize (0.89-0.99)} & 0.94 {\scriptsize (0.82-1.00)} \\
            NVD  & 0.82 {\scriptsize (0.74-0.89)} & 0.58 {\scriptsize (0.48-0.68)} & 0.62 {\scriptsize (0.49-0.74)} & 0.64 {\scriptsize (0.53-0.74)} & 0.95 {\scriptsize (0.88-1.00)} & 0.94 {\scriptsize (0.83-1.00)} \\
            \hline
            ShapeDNA + OC-SVM & 0.66 {\scriptsize (0.56-0.74)} & 0.52 {\scriptsize (0.42-0.60)} & 0.66 {\scriptsize (0.55-0.76)} & 0.47 {\scriptsize (0.35-0.60)} & 0.80 {\scriptsize (0.69-0.90)} & 0.90 {\scriptsize (0.81-0.98)} \\
            LFV + OC-SVM & 0.93 {\scriptsize (0.89-0.97)} & 0.58 {\scriptsize (0.49-0.66)} & 0.67 {\scriptsize (0.57-0.76)} & 0.65 {\scriptsize (0.53-0.77)} & 0.92 {\scriptsize (0.88-0.96)} & 0.98 {\scriptsize (0.95-1.00)} \\
            \hline
            LeNet-CAE + OC-SVM & 0.95 {\scriptsize (0.91-0.98)} & 0.48 {\scriptsize (0.38-0.58)} & 0.50 {\scriptsize (0.37-0.61)} & 0.46 {\scriptsize (0.40-0.51)} & 0.77 {\scriptsize (0.67-0.86)} & 0.69 {\scriptsize (0.56-0.81)}  \\
            Oktay \textit{et al.} + OC-SVM & 0.94 {\scriptsize (0.90-0.98)} & 0.45 {\scriptsize (0.35-0.56)} & 0.48 {\scriptsize (0.35-0.59)} & 0.47 {\scriptsize (0.42-0.53)} & 0.80 {\scriptsize (0.69-0.88)} & 0.76 {\scriptsize (0.65-0.86)} \\
            \hline
            AD classification & 0.49 {\scriptsize (0.47-0.50)} & 0.64 {\scriptsize (0.56-0.70)} & 0.73 {\scriptsize (0.63-0.83)} & {\scriptsize (Used for training)} & 0.52 {\scriptsize (0.48-0.56)} & 0.54 {\scriptsize (0.48-0.61)}  \\
            \hline
            \textbf{Deep NORAH (w/o CAE pretr.)} & 0.79 {\scriptsize (0.71-0.87)} & 0.59 {\scriptsize (0.48-0.69)} & 0.99 {\scriptsize (0.99-1.00)} & 0.76 {\scriptsize (0.67-0.84)} & {\bfseries1.00} {\scriptsize (0.99-1.00)} & {\bfseries1.00} {\scriptsize (0.99-1.00)} \\
            \textbf{Deep NORAH (w/o FC dim. red.)} & 0.98 {\scriptsize (0.94-1.00)} & 0.70 {\scriptsize (0.60-0.79)} & 0.76 {\scriptsize (0.65-0.87)} & 0.65 {\scriptsize (0.54-0.76)} & {\bfseries1.00} {\scriptsize (1.00-1.00)} & {\bfseries1.00} {\scriptsize (1.00-1.00)} \\
            \textbf{Deep NORAH (with CAE pretr.)} & {\bfseries0.99} {\scriptsize (0.99-1.00)} & {\bfseries0.93} {\scriptsize (0.87-0.97)} & {\bfseries1.00} {\scriptsize (0.99-1.00)} & {\bfseries0.92} {\scriptsize (0.86-0.96)} & {\bfseries1.00} {\scriptsize (1.00-1.00)} & {\bfseries1.00} {\scriptsize (1.00-1.00)} \\
            \hline
            \end{tabular}}
\end{table}

\subsubsection{Ablation analysis.} 
\begin{figure}[t!]
\begin{floatrow}
\ffigbox[0.28\textwidth]{%
  \centering
    \includegraphics[width=0.85\columnwidth]{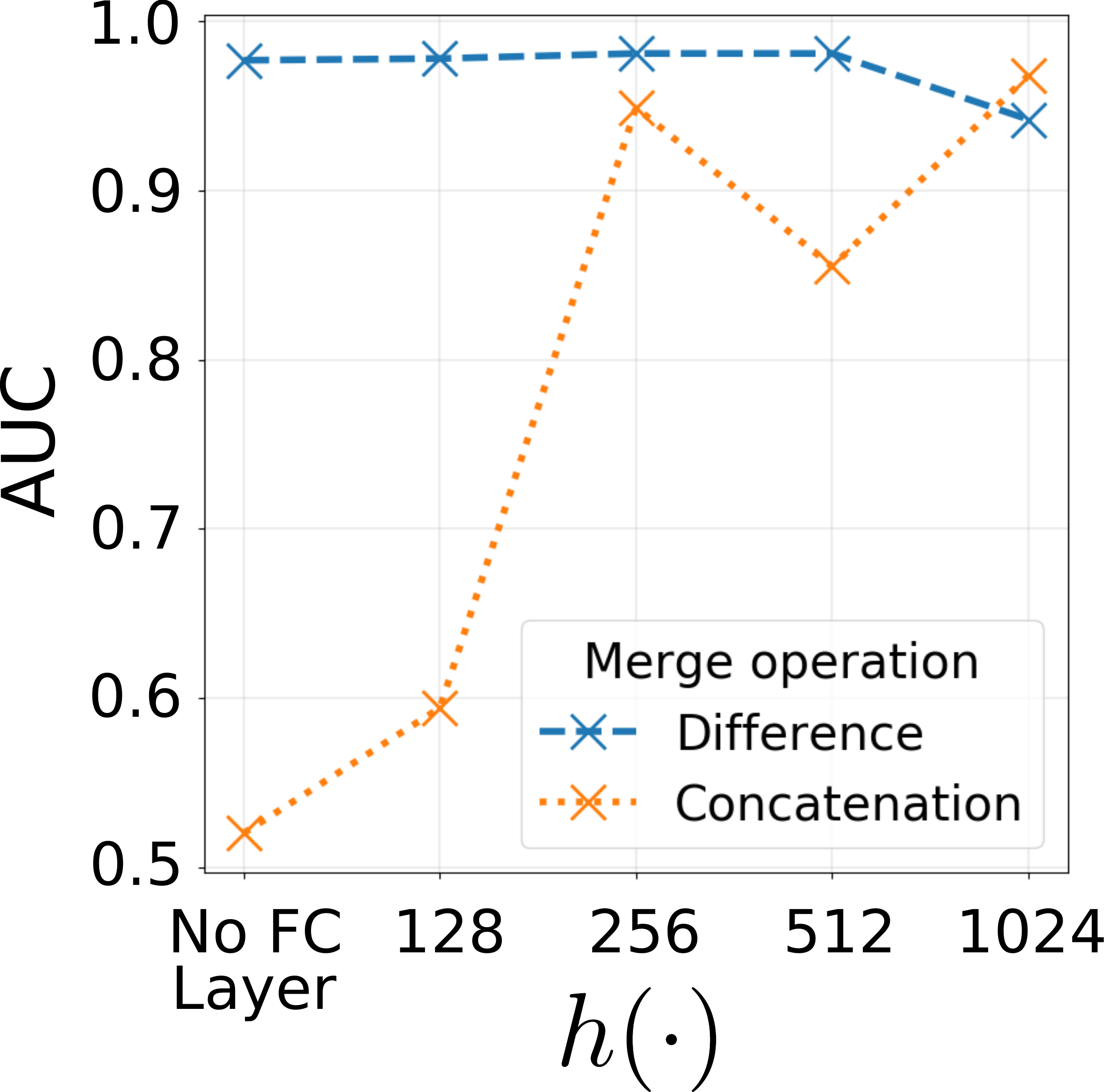}%
}{%
  \caption{APH ablation.}
	\label{fig:ablation-APH}
}
\capbtabbox{%
\resizebox{0.70\columnwidth}{!}{
        \begin{tabular}{C{2cm}|c|C{1.5cm}|C{1.5cm}|C{1.5cm}||C{1.5cm}}
            \hline
            Backbone architecture & Siamese & Sagittal flip & Merge operation & Num. params. & AUC ($S_{\sigma=\text{all}}$)  \\
            \hline
            LeNet & \xmark & \xmark & N/A & 5.4M & 0.62\\
            LeNet & \cmark & \xmark & Concat. & 2.7M & 0.71 \\
            LeNet & \cmark & \xmark & Diff. & 2.7M & 0.76 \\
            LeNet & \cmark & \cmark & Concat. & 2.7M & 0.73 \\
            LeNet & \cmark & \cmark & Diff. & 2.7M & {\bfseries0.95} \\
            \hline
            Deeper & \xmark & \xmark & N/A & 1.1M & 0.58 \\
            Deeper & \cmark & \xmark & Concat. & 555K & 0.72 \\
            Deeper & \cmark & \xmark & Diff. & 555K & 0.81 \\
            Deeper & \cmark & \cmark & Concat. & 555K & 0.75 \\
            Deeper & \cmark & \cmark & Diff. & 555K & 0.94 \\
            \hline
            \end{tabular}
}
}{%
\caption{Shape encoder ablation.}%
\label{table:ablation-encoder}
}
\end{floatrow}

\end{figure}

Table~\ref{AUC table} includes results with and without our CAE pretraining. This stage significantly improve performance in synthetic, MCI and AD (OASIS) cases, perhaps due to a better estimate of $\mathbf{c}$. Fig.~\ref{fig:ablation-APH} illustrates the variations in AUC in $S_{\sigma=\text{all}}$ when changing the merge operation and the size of $h(\cdot)$. Using differences seems to be much more efficient in terms of capacity usage, with almost the same AUC obtained without a FC layer for dimensionality reduction (0.977) and with a FC layer with 512 outputs (0.997). Nevertheless, when applied on real cases, adding this additional component aids to improve the discrimination performance (see Table~\ref{AUC table}). Finally, Table~\ref{table:ablation-encoder} shows results in $S_{\sigma=\text{all}}$ for different encoder settings. Using a Siamese approach with difference as merge operation was in all cases superior than the other alternatives. In terms of architecture, LeNet and deep backbones showed similar performance, with LeNet being slightly better. This might be due to a higher number of FC parameters, as the output of the shape encoder is a vector with $>$32k features.

\newfloatcommand{capbtabbox}{table}[][\FBwidth]

\section{Conclusions} 
\label{sec:conclusions}

We presented a novel anomaly detection-based method for automatically characterizing normal asymmetry in homologous brain structures.
Supervised alternatives restrict the definition of normal individuals due to explicitly learning their differences with respect to subjects with a specific condition. 
This implies that they ignore the asymmetry in control subjects, capturing only the asymmetries induced by the analyzed disease~\cite{liu2019using}, and requiring retraining to detect new conditions unseen during training.
Conversely, our unsupervised alternative leverages a recently introduced one-class-based objective to learn the space of normal asymmetries. 
Hence, it can detect diseased samples by quantifying their distance to the control space. 
Our experiments on hippocampus data showed that our approach could effectively use symmetry information to characterize normal populations and then identify disease presence by contrast, even though only NC subjects are used for training. Our model can potentially be applied to other homologous brain structures and diverse cohorts to aid radiologists in quantifying asymmetries of a normal brain better. In its current form, the model inherits the limitations of the segmentation approach, although it showed to achieve good performance using the outputs of HippMapp3r. Furthermore, it has the burden of not offering qualitative feedback, so future work should focus on bringing interpretability to this tool, e.g., by means of occlusion analysis.

\subsubsection*{Acknowledgments.}
This study was funded by PIP GI 2021-2023 0102472 (CONICET) and PICTs 2019-00070, 2020-00045 and 2021-00023 (Agencia I+D+i).


\bibliographystyle{splncs04}
\bibliography{mybibliography}

\begin{thebibliography}{10}
\providecommand{\url}[1]{\texttt{#1}}
\providecommand{\urlprefix}{URL }
\providecommand{\doi}[1]{https://doi.org/#1}

\bibitem{ADNI}
{ADNI}: Alzheimer’s disease neuroimaging initiative.
  \url{http://adni.loni.usc.edu/}, [Online; accessed February 9th 2023]

\bibitem{IXI}
{IXI} dataset website. \url{http://brain-development.org/ixidataset/}, [Online;
  accessed February 9th 2023]

\bibitem{ardekani2019sexual}
Ardekani, B., et~al.: Sexual dimorphism and hemispheric asymmetry of
  hippocampal volumetric integrity in normal aging and alzheimer disease.
  American Journal of Neuroradiology  \textbf{40}(2),  276--282 (2019)

\bibitem{bernasconi2003mesial}
Bernasconi, N., et~al.: Mesial temporal damage in temporal lobe epilepsy: a
  volumetric {MRI} study of the hippocampus, amygdala and parahippocampal
  region. Brain  \textbf{126}(2),  462--469 (2003)

\bibitem{borchert2021artificial}
Borchert, R., et~al.: Artificial intelligence for diagnosis and prognosis in
  neuroimaging for dementia; a systematic review. medRxiv pp. 2021--12 (2021)

\bibitem{csernansky2004abnormalities}
Csernansky, J.G., et~al.: Abnormalities of thalamic volume and shape in
  schizophrenia. American Journal of Psychiatry  \textbf{161}(5),  896--902
  (2004)

\bibitem{fu2021altered}
Fu, Z., et~al.: Altered neuroanatomical asymmetries of subcortical structures
  in subjective cognitive decline, amnestic mild cognitive impairment, and
  alzheimer’s disease. Journal of Alzheimer's Disease  \textbf{79}(3),
  1121--1132 (2021)

\bibitem{goubran2020hippocampal}
Goubran, M., et~al.: Hippocampal segmentation for brains with extensive atrophy
  using three-dimensional convolutional neural networks. Tech. rep., Wiley
  Online Library (2020)

\bibitem{herbert2005brain}
Herbert, M.R., et~al.: Brain asymmetries in autism and developmental language
  disorder: a nested whole-brain analysis. Brain  \textbf{128}(1),  213--226
  (2005)

\bibitem{herzog2021brain}
Herzog, N.J., Magoulas, G.D.: Brain asymmetry detection and machine learning
  classification for diagnosis of early dementia. Sensors  \textbf{21}(3), ~778
  (2021)

\bibitem{li2021hippocampal}
Li, A., et~al.: Hippocampal shape and asymmetry analysis by cascaded
  convolutional neural networks for alzheimer’s disease diagnosis. Brain
  Imaging and Behavior pp. 1--10 (2021)

\bibitem{liu2019using}
Liu, C.F., et~al.: Using deep siamese neural networks for detection of brain
  asymmetries associated with {Alzheimer's} disease and mild cognitive
  impairment. Magnetic resonance imaging  \textbf{64},  190--199 (2019)

\bibitem{low2019asymmetrical}
Low, A., et~al.: Asymmetrical atrophy of thalamic subnuclei in {Alzheimer's}
  disease and amyloid-positive mild cognitive impairment is associated with key
  clinical features. Alzheimer's \& Dementia: Diagnosis, Assessment \& Disease
  Monitoring  \textbf{11}(1),  690--699 (2019)

\bibitem{marcus2010open}
Marcus, D.S., et~al.: Open access series of imaging studies: longitudinal
  {{MRI}} data in nondemented and demented older adults. Journal of cognitive
  neuroscience  \textbf{22}(12),  2677--2684 (2010)

\bibitem{oktay2017anatomically}
Oktay, O., et~al.: Anatomically constrained neural networks (acnns):
  application to cardiac image enhancement and segmentation. IEEE transactions
  on medical imaging  \textbf{37}(2),  384--395 (2017)

\bibitem{andrade2015defining}
de~Oliveira, A., et~al.: Defining multivariate normative rules for healthy
  aging using neuroimaging and machine learning: an application to
  {Alzheimer's} disease. Journal of Alzheimer's Disease  \textbf{43}(1),
  201--212 (2015)

\bibitem{park2022topographic}
Park, B.y., et~al.: Topographic divergence of atypical cortical asymmetry and
  atrophy patterns in temporal lobe epilepsy. Brain  \textbf{145}(4),
  1285--1298 (2022)

\bibitem{pedraza2004asymmetry}
Pedraza, O., Bowers, D., Gilmore, R.: Asymmetry of the hippocampus and amygdala
  in {MRI} volumetric measurements of normal adults. JINS  \textbf{10}(5),
  664--678 (2004)

\bibitem{penny2011statistical}
Penny, W.D., et~al.: Statistical parametric mapping: the analysis of functional
  brain images. Elsevier (2011)

\bibitem{princich2021diagnostic}
Princich, J.P., et~al.: Diagnostic performance of {MRI} volumetry in epilepsy
  patients with hippocampal sclerosis supported through a random forest
  automatic classification algorithm. Frontiers in Neurology  \textbf{12},
  613967 (2021)

\bibitem{richards2020increased}
Richards, R., et~al.: Increased hippocampal shape asymmetry and volumetric
  ventricular asymmetry in autism spectrum disorder. NeuroImage: Clinical
  \textbf{26},  102207 (2020)

\bibitem{ruff2018deep}
Ruff, L., et~al.: Deep one-class classification. In: ICML. pp. 4393--4402. PMLR
  (2018)

\bibitem{scholkopf1999support}
Sch{\"o}lkopf, B., et~al.: Support vector method for novelty detection. NIPS
  \textbf{12} (1999)

\bibitem{tortora2018principles}
Tortora, G.J., Derrickson, B.H.: Principles of anatomy and physiology. John
  Wiley \& Sons (2018)

\bibitem{van2021elasticdeform}
van Tulder, G.: elasticdeform: Elastic deformations for n-dimensional images
  (2021)

\bibitem{wachinger2016whole}
Wachinger, C., et~al.: Whole-brain analysis reveals increased neuroanatomical
  asymmetries in dementia for hippocampus and amygdala. Brain
  \textbf{139}(12),  3253--3266 (2016)

\bibitem{woolard2012anatomical}
Woolard, A.A., Heckers, S.: Anatomical and functional correlates of human
  hippocampal volume asymmetry. Psychiatry Research: Neuroimaging
  \textbf{201}(1),  48--53 (2012)

\bibitem{zhang2021anomaly}
Zhang, Z., Deng, X.: Anomaly detection using improved deep {SVDD} model with
  data structure preservation. Pattern Recognition Letters  \textbf{148}, ~1--6
  (2021)

\end{thebibliography}

\end{document}